\documentclass[fleqn,twoside]{article}
\usepackage{espcrc2}

\usepackage{graphicx}
\usepackage[figuresright]{rotating}

\newcommand{\AmS}{{\protect\the\textfont2
  A\kern-.1667em\lower.5ex\hbox{M}\kern-.125emS}}

\title{Quantum Markov Process on a Lattice}

\author{T. Hashimoto, M. Horibe and A. Hayashi \\\ \\
        Department of Applied Physics, Fukui University, Bunkyo 3-9-1, Fukui 910-8507, Japan}
       
\begin{document}

\begin{abstract}
We develop a systematic description of Weyl and Fano operators
on a lattice phase space.
Introducing the so-called ghost variable even on an odd lattice,
odd and even lattices can be treated in a symmetric way.
The Wigner function is defined using these operators on the quantum
phase space, which can be interpreted as a spin phase space.
If we extend the space with a dichotomic variable,
a positive distribution function can be defined on the new space.
It is shown that there exits a quantum Markov process on the extended
space which describes the time evolution of the distribution function.
\vspace{1pc}
\end{abstract}

\maketitle

%%%%%%%%%%%%%%
% Section 1. %
%%%%%%%%%%%%%%

\section{
Introduction
}

The quantum Markov process for an integer spin system
was constructed on the lattice phase space by Cohendet
et al. \cite{CCSS(1988)}.
The time evolution of the
pseudo-distribution function, i.e., the Wigner function
on the space can be derived from the process.
However, the extension of their method to a half integer spin system
is not straightforward.
There are difficulties in the construction of the Wigner function on an even lattice.
A way to avoid this difficulties is to introduce the so-called "ghost variable"
\cite{Leonhardt}.
The Weyl and Fano operators can be constructed
on an even lattice if we consider fictitious lattice points between
the original ones (see also \cite{Ours}).
In the present paper, we develop the systematic treatment of
the Wigner function on both odd and even lattices and
show that a quantum Markov process for a half integer spin system
can also be constructed in a similar manner.

%%%%%%%%%%%%%%
% Section 2. %
%%%%%%%%%%%%%%

\section{
Weyl and Fano operators with ghost variables on a lattice
}

We consider a lattice composed of \(N\) lattice points.
A lattice with odd or even lattice points is called odd or even lattice,
respectively.
We denote the set of integers and half integers by \(\bar{\bf Z}\).
For odd \(N\), we define \(J^o, {\bar J}^o\) as the sets composed of integers, integers and half integers
between \(-(N/2)\) and \((N/2)+1\),
respectively, and \({\bar J}^o_+\) be the non-negative part of \({\bar J}^o\).
For even \(N\), we define \(J^e, {\bar J}^e, {\bar J}^e_+\) similarly
between \(-(N/2)\) and \((N-1)/2\).
The symbols \(J, {\bar J}, {\bar J}_+\) are
the abbreviations of \(J^o, {\bar J}^o, {\bar J}^o_+\) for odd \(N\) and
\(J^e, {\bar J}^e, {\bar J}^e_+\) for even \(N\).
We consider a Kronecker's delta function \(\bar{\delta}^{(N)}_{n,m}\) on \(\bar{\bf Z}\),
defined by \(\bar{\delta}^{(N)}_{m,n}\equiv\delta_{2m,2n}\) in mod \(2N\),
where \(m,n\in\bar{\bf Z}\).
We denote one of primitive roots of unity of order \(N\) by \(\omega\), e.g.,
\(\omega=\exp\left({2\pi i\over N}\right)\).

The \((n,m)\) components of basic phase, shift and skew operators \(Q, P, T\) are defined as

\begin{equation}
Q\equiv(\omega^n\bar{\delta}^{(N)}_{n,m}),
P\equiv(\bar{\delta}^{(N)}_{n+1,m}),
T\equiv(\bar{\delta}^{(N)}_{n+m,0}),
\end{equation}

\noindent
for odd \(N\), and

\[
Q\equiv(\omega^n\bar{\delta}^{(N)}_{n,m}),
P\equiv((1-2\bar{\delta}^{(N)}_{n,{N-1\over2}})\bar{\delta}^{(N)}_{n+1,m}),
\]
\begin{equation}
T\equiv(\bar{\delta}^{(N)}_{n+m,0}),
\end{equation}

\noindent
for even \(N\), where \(m,n \in J\).
The Froquet or Bloch angle is \(\pi\) in the latter case \cite{RA(1999)}.

%%%%%%%%%%%%%%
% Section 3. %
%%%%%%%%%%%%%%

\section{
Wigner function on a lattice
}

\subsection{Weyl operators}

We define the Weyl operators by

\begin{equation}
W_{m,n}\equiv\omega^{-2mn}Q^{2n}P^{-2m}
=\omega^{2mn}P^{-2m}Q^{2n},
\end{equation}

\noindent
for both odd and even \(N\), where \(m, n\in \bar{J}\).
We can see the Weyl operators satisfy the relations for reflection
and shift:

\begin{equation}
W^\dagger_{m,n}=W_{-m,-n}=TW_{m,n}T,
\end{equation}

\begin{equation}
W_{m,n}W_{m',n'}=\omega^{-2(mn'-nm')}W_{m+m',n+n'}.
\end{equation}

The \(N^2\) operators \(W_{m,n}\) with \((m, n)\) in the basic region
\(\bar{J}_+\times\bar{J}_+\) are complete and orthogonal in the trace norm.
Generally, \(W_{m,n}\) with \((m,n)\in\bar{\bf Z}\times\bar{\bf Z}\) and 
\(W_{{\rm mod}(m,{N\over2}),{\rm mod}(n,{N\over2})}\) in the basic region are
related as

\begin{equation}
W_{m,n}=\theta(m,n)W_{{\rm mod}(m,{N\over2}),{\rm mod}(n,{N\over2})},
\end{equation}

\noindent
where \( \theta(m,n)=(-1)^{\tau(m,n)}, \) and

\begin{equation}
\tau(m,n)=2n\left[{2m\over N}\right]+2m\left[{2n\over N}\right]
+\left[{2m\over N}\right]\cdot\left[{2n\over N}\right],
\end{equation}

\noindent
for odd \(N\),

\begin{equation}
\tau(m,n)=2n\left[{2m\over N}\right]+2m\left[{2n\over N}\right]
+\left[{2m\over N}\right]+\left[{2n\over N}\right],
\end{equation}

\noindent
for even \(N\).

The Weyl operators have the next symplectic property,
i.e., they are rotated by \(\pi/2\) anti-clockwise under
the Fourier transformation,

\begin{equation}
{\cal F}W_{m,n}{\cal F}^\dagger=W_{-n,m},\ {\cal F}={1\over\sqrt{N}}(\omega^{mn}).
\end{equation}

\subsection{Fano operators}

We define the Fano operators by

\begin{equation}
\Delta_{m,n}=W_{m,n}T=\omega^{-2mn}Q^{2n}TP^{2m},
\end{equation}

\noindent
for both odd and even \(N\), where \(m, n \in \bar{J}\).
They are hermite but over complete.
Those in the basic region are orthogonal and complete.
The Fano operators have the same symplectic property as
the Weyl operators,

\begin{equation}
{\cal F}\Delta_{m,n}{\cal F}^\dagger=\Delta_{-n,m}.
\end{equation}

We can see that the Fano operators have proper marginal properties:

\begin{equation}
\sum_{n\in\bar{J}}\Delta_{m,n}=0, \ {\rm for\ half\ integer\ } m \hspace{2mm}{\rm (ghost)}
\end{equation}

\begin{equation}
\sum_{n\in\bar{J}}\Delta_{m,n}=2N(\delta^{(N)}_{i,m}\delta^{(N)}_{j,m}), \ {\rm for\ integer\ } m
\end{equation}

\begin{equation}
\sum_{m\in\bar{J}}\Delta_{m,n}=0, \ {\rm for\ half\ integer\ } n \hspace{2mm}{\rm (ghost)}
\end{equation}

\begin{equation}
\sum_{m\in\bar{J}}\Delta_{m,n}=2(\omega^{m(i-j)}), \ {\rm for\ integer\ } n
\end{equation}

\noindent
for odd \(N\), and

\begin{equation}
\sum_{n\in\bar{J}}\Delta_{m,n}=0, \ {\rm for\ integer\ } m \hspace{2mm}{\rm (ghost)}
\end{equation}

\begin{equation}
\sum_{n\in\bar{J}}\Delta_{m,n}=2N(\delta^{(N)}_{i,m}\delta^{(N)}_{j,m}),{\rm for\ half\ integer\ } m
\end{equation}

\begin{equation}
\sum_{m\in\bar{J}}\Delta_{m,n}=0, \ {\rm for\ integer\ } n \hspace{2mm}{\rm (ghost)}
\end{equation}

\begin{equation}
\sum_{m\in\bar{J}}\Delta_{m,n}=2(\omega^{n(i-j)}), \ {\rm for\ half\ integer\ } n
\end{equation}

\noindent
for even \(N\).

\subsection{Time evolution of the Wigner function}

We define the Wigner function by

\begin{equation}
{\cal W}(m,n)={1\over N}{\rm tr}(\rho\Delta_{m,n}),
\end{equation}

\noindent
which is real valued and bounded by \(1/N\),

\begin{equation}
\label{wigner_bound}
-{1\over N}\le {\cal W}(m,n)\le{1\over N}.
\end{equation}

\noindent
The average value of an observable \({\cal O}\) is given by

\begin{equation}
\langle{\cal O}\rangle = N\sum_{m,n\in\bar{J}_+}{\cal O}(m,n){\cal W}(m,n),
\end{equation}

\noindent
where \({\cal O}(m,n)={1\over N}{\rm tr}(O\Delta_{m,n})\).
We rewrite the time evolution equation of the density matrix \(\rho\),

\begin{equation}
i{\partial\rho_t\over\partial t}=[H,\rho_t],
\end{equation}

\noindent
in terms of the Wigner function.
We expand the Hamiltonian \(H\) by using the Weyl operators in the
basic region,
and the density matrix by the Fano operators in the
same region.
\noindent
Equating the coefficient of \(\Delta_{m,n}\), the time evolution equation
of the Wigner function is given by

\[
{d\over dt}{\cal W}_{mn}=-2
\sum_{m'n'\in \bar{J}_+}
\tilde{\cal H}^{+}(m'',n'')
\]
\[
\sin\left\{{{2\pi i\over N}\{2(m''n'-n''m')-\alpha(m'',n'')\}}
\right\}
\]
\begin{equation}
\theta(m''+m',n''+n')
{\cal W}_{m'n'},
\end{equation}

\noindent
where
\(m''={\rm mod}(m-m',{N\over2})\),
\(n''={\rm mod}(n-n',{N\over2})\)
and
\(\tilde{\cal H}^+(m,n)\) and \(\alpha(m,n)\) are determined by
the polar decomposition
\(\tilde{\cal H}(m,n)=\tilde{\cal H}^+(m,n)\omega^{\alpha(m,n)}\).

%%%%%%%%%%%%%%
% Section 4. %
%%%%%%%%%%%%%%

\section{
Construction of a Markov process 
}

%%% the(?) dichotomic variable
We extend the basic region with the dichotomic variable
\(\sigma\in\{\pm1\}\equiv B\), i.e.,
\( \bar{J}_+\times\bar{J}_{+}\rightarrow
\bar{J}_+\times\bar{J}_{+}\times B \), and
consider a new real valued function \(G(m,n,\sigma)\) on the extended space,

\begin{equation}
G(m,n,\sigma)={1\over4N}\left\{{2\over N}+\sigma{\cal W}(m,n)\right\}.
\end{equation}

\noindent
It can be seen that \(G(m,n,\sigma)\) is positive and satisfies the inequality, \\

\begin{equation}
{1\over4N^2}\le G(m,n,\sigma)\le{3\over4N^2},
\end{equation}

\noindent
and normalization condition,

\begin{equation}
\sum_{(m,n)\in\bar{J}_+\times\bar{J}_{+},\ \sigma\in\{\pm1\}}G(m,n,\sigma)=1.
\end{equation}

\noindent
The positivity of \(G(m,n,\sigma)\) follows from the boundedness of
the Wigner function Eq.(\ref{wigner_bound}).
The average value of an observable \({\cal O}\) is given by

\begin{equation}
\langle{\cal O}\rangle = 2N^2\sum_{m,n\in\bar{J}_+, \sigma\in B}\sigma{\cal O}(m,n)G(m,n,\sigma).
\end{equation}

\noindent
It is natural to call \(G(m,n,\sigma)\) as a distribution function on
the extended space \(\bar{J}_+\times\bar{J}_{+}\times B \).
The time evolution of \(G(m,n,\sigma)\) is given by

\[
{d\over dt}G(m,n,\sigma)
=\sum_{m'n'\in \bar{J}_+, \sigma'\in\{\pm1\}}
-{\rm sgn}(\sigma\sigma')
\]

\[
\tilde{\cal H}^+(m'',n'')
\sin\left\{{{2\pi i\over N}\{2(m''n'-n''m')-\alpha(m'',n'')\}}\right\}
\]

\begin{equation}
\times\theta(m''+m',n''+n')
G(m',n',\sigma').
\end{equation}

\noindent
Introducing a generating operator as

\[ {\cal A}_t(m,n,\sigma;m',n',\sigma')
=
\tilde{\cal H}^+(m'',n'')
\left[{1\over G(m',n',\sigma')}
\right.
\]
\[
-{\rm sgn}(\sigma\sigma')
\times\sin\left\{{{2\pi i\over N}
\{2(m''n'-n''m')-\alpha(m'',n'')\}}\right\}
\]
\begin{equation}
\left.
\theta(m''+m',n''+n')\right],
\end{equation}

\noindent
for \((m,n)\ne(m',n')\),

\begin{equation}
{\cal A}_t(m,n,\sigma;m,n,\sigma')=0,
\end{equation}

\noindent
for \((m,n)=(m',n'),\ \sigma\ne\sigma'\),

\[
{\cal A}_t(m,n,\sigma;m,n,\sigma)=-{2\over G(m,n,\sigma)}
\]
\begin{equation}
\sum_{(m',n')\in\bar{J}_+\times\bar{J}_{+}\backslash(0,0)}\tilde{\cal H}^+(m',n')
\end{equation}

\noindent
for \((m,n,\sigma)=(m',n',\sigma')\),
the equation for \(G(m,n,\sigma)\) can be written briefly as

\[
{d\over dt}G(m,n,\sigma)=
\sum_{m',n'\in\bar{J}_+,\ \sigma\in\{\pm1\}}
\]
\begin{equation}
{\cal A}_t(m,n,\sigma;m',n',\sigma')G(m',n',\sigma').
\end{equation}

\noindent
It can be checked that the generator \({\cal A}_t(m,n,\sigma;m',n',\sigma')\)
satisfies the Markov condition,

\begin{equation}
\sum_{(m,n,\sigma)\in\bar{J}_+\times\bar{J}_+\times B}{\cal A}(m,n,\sigma;m',n',\sigma')=0,
\end{equation}

\begin{equation}
{\cal A}(m,n,\sigma;m',n',\sigma')\ge0, \ {\rm if}\ (m,n,\sigma)\ne(m',n,\sigma'),
\end{equation}

\noindent
which assures the existence of a background quantum Markov process
on the extended lattice quantum phase space.

%%%%%%%%%%%%%%
% Referecnes %
%%%%%%%%%%%%%%

\end{document}